\documentclass[12pt]{article}
\usepackage{amsmath}
\usepackage{amssymb}
\usepackage{graphicx}
\usepackage{float}
\usepackage{indentfirst}
\usepackage{mathrsfs}

\usepackage{setspace}

\usepackage[labelfont=bf,labelsep=period]{caption}

\usepackage[numbers,sort&compress]{natbib}

\newtheorem{theorem}{Theorem}

\newtheorem{corollary}{Corollary}
\newtheorem{lemma}{Lemma}

\title{Entanglement Classification of Four-partite States under the SLOCC}

\author
{S. M. Zangi$^1$, Jun-Li Li$^{1,3}$, and Cong-Feng Qiao$^{1,2,3\ast}$\\ [0.2cm]
\normalsize{$^1$Department of Physics, University of the Chinese Academy of Sciences,}\\
\normalsize{YuQuan Road 19A, Beijing 100049, China}\\[2pt]
\normalsize{$^2$ Department of Physics \& Astronomy, York University, Toronto, ON M3J 1P3, Canada}\\[2pt]
\normalsize{$^3$Key Laboratory of Vacuum Physics, University of Chinese Academy of Sciences}\\[3mm]
\normalsize{$^\ast$ To whom correspondence should be addressed; E-mail: qiaocf@ucas.ac.cn.}
}

\date{}

\begin{document}
\baselineskip24pt \maketitle

\begin{abstract}\doublespacing
We present a practical classification scheme for the four-partite entangled states under stochastic local operations and classical communication (SLOCC). By transforming a four-partite state into a triple-state set composed of two tripartite and one bipartite states, the entanglement classification is reduced to the classification of tripartite and bipartite entanglements. This reduction method has the merit of involving only the linear constrains, and meanwhile providing an insight into the entanglement character of the subsystems.
\end{abstract}

\section{Introduction}

Entanglement, the peculiar feature and marked difference of quantum theory from classical ones, is now regarded as the main physical resource in quantum information sciences \cite{QCQI}. By dint of entanglement, various counterintuitive and unique applications are emerging, e.g. quantum teleportation \cite{Teleportation}, super dense coding \cite{dense-coding92,dense-coding96}, and quantum cryptography protocols \cite{crypto-bell}, etc. Since two states belonging to the same equivalent class under SLOCC may be employed to perform the same quantum information tasks \cite{three-qubit}, the entanglement classification plays an important role in quantum information theory. Although having been intensively studied, we still have a very limited knowledge on the general entanglement classes under SLOCC for systems of more complex than four-qubit entangled states.

There exist only two different kinds of genuine tripartite entangled states in pure three-qubit systems under SLOCC \cite{three-qubit}, i.e. the GHZ and W states. The number of entanglement classes increases dramatically with the increase of particles and dimensions of the entangled state. It turns out that the number of classes for a general four-qubit system is infinite, in nine different entanglement families \cite{four-qubit-nine}. When more particles are involved, the existing operational classification schemes are only applicable to the high symmetric states \cite{N-symmetric}. For a general multipartite pure state, the coefficient matrix method can only identify the discrete entanglement families with different ranks, which is a rather coarse grain classification per se \cite{N-Coefficient-M}. It is shown that the geometric relations of individual particles are capable of characterizing the different entanglement classes of multipartite state under the SLOCC \cite{Polytopes}. Moreover, the algebraic invariants were explored to distinguish the entanglement classes \cite{Infinite-SLOCC}, where a complete set of invariants usually involves some complicated expressions and the individuality of each particle is not explicitly manifested. Despite these progresses, a practical method of verifying the SLOCC equivalence of two arbitrarily given multipartite states is highly desirable. More importantly, it is still unclear, for a multipartite entangled state, how entanglement characters of subsystems behave and generate the whole nature.

In this paper we present a general classification scheme for four-partite pure states of arbitrary but finite dimensions. By applying singular value decomposition to a bipartition of the system, a four-partite state is then transformed into a triple-state set composed of two tripartite states and one bipartite state. And the two four-partite quantum states are thought to be SLOCC equivalent if and only if the quantum states in the the triple-state sets are SLOCC equivalent respectively. Our method provides a systematic procedure to reduce the entanglement classification of multipartite states to that of less partite states, and hence to distinguish the entanglement classes of the whole system through its subsystems.

\section{The classification scheme}
\subsection{The representations of quantum states}

A pure one particle quantum state may be represented by a normalized complex vector in Hilbert space, while a bipartite state of $|\Psi_{I_1I_2}\rangle = \sum_{i_1,i_2=1}^{I_1,I_2} \psi_{i_1i_2}|i_1\rangle|i_2\rangle$ may be expressed in matrix form
\begin{eqnarray}
\Psi_{I_1I_2} =
\begin{pmatrix}
\psi_{11} & \psi_{12} & \cdots & \psi_{1I_2} \\
\psi_{21} & \psi_{22} & \cdots & \psi_{2I_2} \\
\vdots & \vdots & \ddots & \vdots \\
\psi_{I_11} & \psi_{I_12} & \cdots & \psi_{I_1I_2}
\end{pmatrix}\; .
\end{eqnarray}
Two $N$-partite states $|\Psi'\rangle$ and $|\Psi\rangle$ are SLOCC equivalent if and only if they can be connected via invertible local operators, i.e. $|\Psi'\rangle = A_1 \otimes \cdots \otimes A_N |\Psi\rangle$ \cite{three-qubit}. For bipartite quantum states in matrix form, the SLOCC equivalence of $|\Psi_{I_1I_2}'\rangle = A_1\otimes A_2 |\Psi_{I_1I_2}\rangle$  turns to
\begin{eqnarray}
\Psi_{I_1I_2}'=A_1\Psi_{I_1I_2}A_2^{\mathrm{T}} \; .
\end{eqnarray}
Here $A_1\in \mathbb{C}^{I_1\times I_1}$ and $A_2\in \mathbb{C}^{I_2\times I_2}$ are invertible matrices, and the superscript $\mathrm{T}$ means the matrix transposition. A tripartite state may be expressed as a tuple of the matrices \cite{2NN,LNN}
\begin{eqnarray}
\Psi_{I_1I_2I_3} = (\Gamma_1,\Gamma_2, \cdots, \Gamma_{\!I_1})\; ,
\end{eqnarray}
where $\Gamma_i \in \mathbb{C}^{I_2\times I_3}$ for $i\in \{1,2,\cdots, I_1\}$. In this case, the SLOCC equivalence of two tripartite states, $|\Psi_{I_1I_2I_3}' \rangle = A_1 \otimes A_2 \otimes A_3 |\Psi_{I_1I_2I_3} \rangle$, may now be expressed as
\begin{eqnarray}
\Psi_{I_1I_2I_3}' & = & (\Gamma'_1,\Gamma'_2, \cdots, \Gamma'_{I_1}) \nonumber \\
&=& (A_2\Gamma_1A_3, A_2\Gamma_2A_3, \cdots, A_2 \Gamma_{I_1}A_3)A_1^{\mathrm{T}} \; .
\end{eqnarray}
Here the tripartite state behaves as a row vector whose components are matrices.

For the representation of four-partite states, we first introduce two operations related to matrices, the vectorization and folding. The vectorization of an $I_1\times I_2$ dimensional matrix $\Psi_{I_1I_2}$ with complex elements $\psi_{ij}$ is
\begin{eqnarray}
\mathcal{V}(\Psi_{I_1I_2}) \equiv
(\psi_{11}, \cdots , \psi_{I_11}, \psi_{12}, \cdots, \psi_{I_12}, \cdots, \psi_{1I_2}, \cdots, \psi_{I_1I_2})^{\mathrm{T}}\; .
\end{eqnarray}
We define the folding operation to be the inverse operation of the vectorization by wrapping a vector into a matrix
\begin{eqnarray}
\mathcal{W}(\vec{a}\,)_{I_1I_2} \equiv
\begin{pmatrix}
a_1 & a_{I_1+1} & \cdots & a_{(I_2-1)\cdot I_1 +1} \\
a_2 & a_{I_1+2} & \cdots & a_{(I_2-1)\cdot I_1 +2} \\
\vdots & \vdots & \ddots & \vdots\\
a_{I_1} & a_{2\cdot I_1} & \cdots & a_{I_1\cdot I_2}
\end{pmatrix} \; , \label{Wrapping-def}
\end{eqnarray}
where $\vec{a} = (a_1,a_2,\cdots, a_{L})^{\mathrm{T}}$, $L=I_1\cdot I_2$. The subscripts $I_1,I_2$ on the left hand side of Eq. (\ref{Wrapping-def}) indicate the dimensions of the obtained matrix which may be omitted when there is no ambiguity in the matrix dimensions.

\begin{figure}[t] \centering
\scalebox{0.6}{\includegraphics{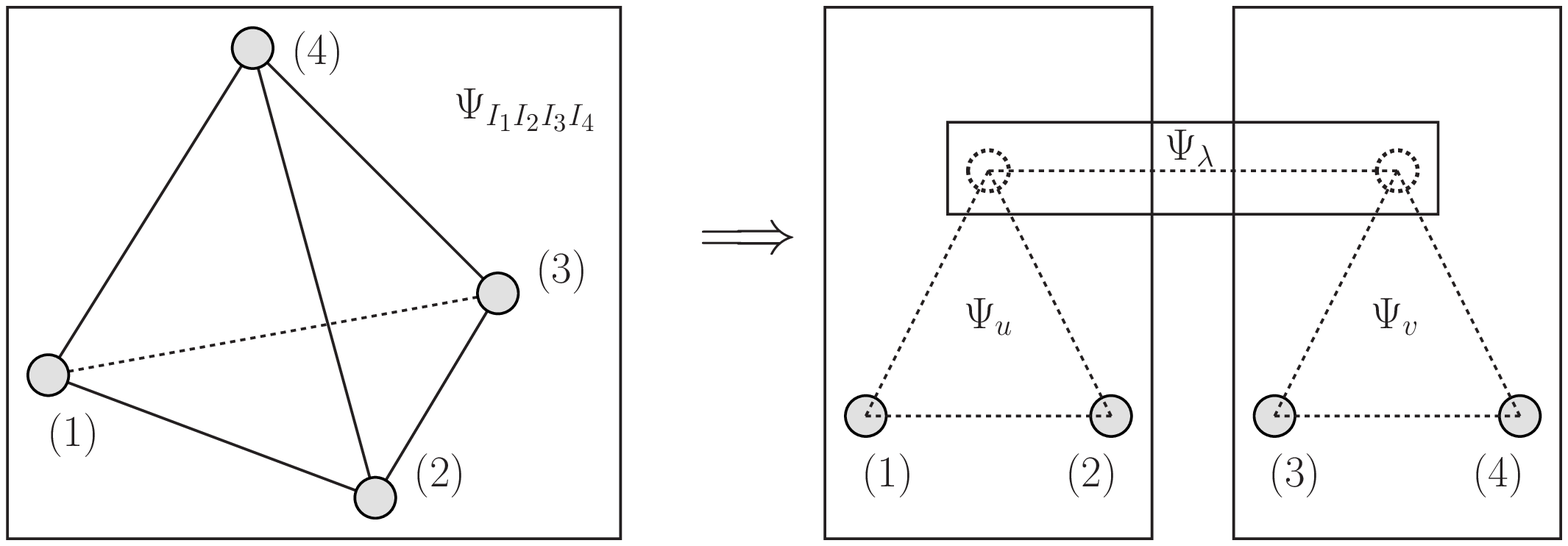}}
\caption{A four-partite state $\Psi_{I_1I_2I_3I_4}$ is factored into a triple-state set according to the bipartition $(I_1I_2)(I_3I_4)$ which includes two tripartite pure state $\Psi_{u}$ and $\Psi_{v}$ and one bipartite pure state $\Psi_{\lambda}$. } \label{fig-1}
\end{figure}

Let $\Psi_{I_1I_2I_3I_4}$ be an $I_1 \times I_2\times I_3\times I_4$ pure quantum state. A bipartition of the four-particle state may be expressed as $\Psi_{(I_1I_2)(I_3I_4)}$, where the four particles are grouped into two composite partitions, i.e.
\begin{eqnarray}
\Psi_{(I_1I_2)(I_3I_4)} =
\begin{pmatrix}
\psi_{1111} & \psi_{1112} & \cdots & \psi_{11I_3I_4} \\
\psi_{1211} & \psi_{1212} & \cdots & \psi_{12I_3I_4} \\
\vdots & \vdots & \ddots & \vdots \\
\psi_{I_1I_211} & \psi_{I_1I_212} & \cdots & \psi_{I_1I_2I_3I_4}
\end{pmatrix}\; . \label{Four-biparite-mat}
\end{eqnarray}
The singular value decomposition of (\ref{Four-biparite-mat}) goes as
\begin{eqnarray}
\Psi_{(I_1I_2)(I_3I_4)} = U \Lambda V^{\dag} \; , \label{Four-bipartite-SVD}
\end{eqnarray}
where $\Lambda$ is a diagonal matrix of rank $r$; the unitary matrices $U$ and $V$ are composed by the left and right singular vectors of $\Psi_{(I_1I_2)(I_3I_4)}$, i.e. $U = (\vec{u}_1,\vec{u}_2,\cdots)$ and $V = (\vec{v}_1,\vec{v}_2,\cdots)$, and the dimensions of vectors $\vec{u}_i$ and $\vec{v}_j$ are $I_1\cdot I_2$ and $I_3\cdot I_4$ respectively; $V^{\dag} = (V^*)^{\mathrm{T}}$ is the conjugate transpose of a matrix. It is legitimate to introduce the following triple-state set expression for the four-partite state based upon the partition $(I_1I_2)(I_3I_4)$:
\begin{eqnarray}
\Psi_{(I_1I_2)(I_3I_4)} = (\Psi_{u},\Psi_{\lambda},\Psi_{v})\; . \label{Four-Rep}
\end{eqnarray}
Here, $\Psi_{u} = (\mathcal{W}(\vec{u}_1), \cdots,\mathcal{W}(\vec{u}_r))$ with $\mathcal{W}(\vec{u_i}) \in \mathbb{C}^{I_1\times I_2}$, $\Psi_{v} = (\mathcal{W}(\vec{v}_1),\cdots, \mathcal{W}(\vec{v}_r))$ with $\mathcal{W}(\vec{v_i}) \in \mathbb{C}^{I_3\times I_4}$, and $\Psi_{\lambda} = \Lambda_r =  \mathrm{diag}\{\lambda_1,\cdots, \lambda_r\}$ with $\lambda_i > 0, i\in \{1,\cdots, r\}$ being the nonzero singular values in $\Lambda$. In this representation, $\Psi_{u}$ and $\Psi_{v}$ may be regarded as the tripartite states of $r\times I_1\times I_2$ and $r\times I_3\times I_4$, and $\Psi_{\lambda}$ is a bipartite state of dimension $r\times r$, as shown in Figure \ref{fig-1}.

In the following we define a complementary state of a tripartite state. If a tripartite state $\Psi_{rI_1I_2} = (\Gamma_1,\cdots, \Gamma_r)$, where $\Gamma_i \in \mathbb{C}^{I_1\times I_2}$ is a genuine entangled state of $r\times I_1\times I_2$, then $\mathcal{V}(\Gamma_i)$, $i\in \{1,\cdots, r\}$ are linearly independent vectors. The complementary state of $\Psi_{rI_1I_2}$ is defined as
\begin{eqnarray}
\overline{\Psi}_{rI_1I_2}\equiv (\Gamma_{r+1},\cdots,\Gamma_{I_1\cdot I_2}) \; .
\end{eqnarray}
Here $\mathcal{V}(\Gamma_{i})$, $i \in \{1,\cdots, r, r+1,\cdots, I_1\cdot I_2\}$, are linearly independent vectors. According to this definition, the complementary state of the $r\times I_1\times I_2$ state $\Psi_{u}$ now can be expressed as
\begin{eqnarray}
\overline{\Psi}_{u} = (\mathcal{W}(\vec{u}_{r+1}), \cdots, \mathcal{W}(\vec{u}_{I_1\cdot I_2})) \; ,
\end{eqnarray}
which is a $(I_1\cdot I_2-r)\times I_1\times I_2$ state with $\mathcal{W}(\vec{u}_i) \in \mathbb{C}^{I_1\times I_2}$. Here the left singular vectors are divided into two parts $U  = (U_1,U_0)$ with $U_1 =  (\vec{u}_1,\cdots,\vec{u}_r)$ and $U_0= (\vec{u}_{r+1}, \cdots, \vec{u}_{I_1\cdot I_2})$. And the quantum state $\overline{\Psi}_{u}$ is obtained by wrapping the left singular vectors that correspond to the singular value zero of $\Psi_{(I_1I_2)(I_3I_4)}$. Similar definitions apply to $\overline{\Psi}_{v}$ as well.

Obviously, for a matrix $A\in \mathbb{C}^{I_1\cdot I_2\times I_1\cdot I_2}$, without loss generality it can be
expressed in the following block-form
\begin{eqnarray}
A = \begin{pmatrix}
A_{11} & A_{12} & \cdots & A_{1I_1} \\
A_{21} & A_{22} & \cdots & A_{2I_1} \\
\vdots & \vdots & \ddots & \vdots   \\
A_{I_11} & A_{I_12} & \cdots & A_{I_1I_1}
\end{pmatrix} \; .
\end{eqnarray}
Here $A_{ij}$ are $I_2\times I_2$ submatrices. The realignment of the matrix
$A$ according to the blocks $A_{ij} \in
\mathbb{C}^{I_2\times I_2}$ is defined to be \cite{Matrix-Realignment}
\begin{eqnarray}
\mathcal{R}(A) \equiv
\begin{pmatrix}
\mathcal{V}(A_{11}), \cdots, \mathcal{V}(A_{I_11}),
\mathcal{V}(A_{12}), \cdots, \mathcal{V}(A_{I_12}),\cdots,\mathcal{V}(A_{I_1I_1})
\end{pmatrix}^{\mathrm{T}} \; , \nonumber
\end{eqnarray}
where $\mathcal{R}(A) \in \mathbb{C}^{I_1\cdot I_1\times I_2\cdot I_2}$.
By means of the complementary state, the following indispensable Lemma exits in following discussion.
\begin{lemma}
Tripartite states $\Psi_{rI_1I_2}'=(\Gamma'_1,\cdots, \Gamma'_r)$ and $\Psi_{rI_1I_2}=(\Gamma_1,\cdots, \Gamma_r)$ are SLOCC equivalent if and only if there exist $\widetilde{P} = \begin{pmatrix}
P & Y \\
0 & \overline{P}
\end{pmatrix}$ such that
\begin{eqnarray}
\mathrm{rank}[\mathcal{R}( U \widetilde{P} U'^{-1})] = 1 \; .
\end{eqnarray}
Here $U'= (U_1',U'_0)$ with $U_1'=(\mathcal{V}(\Gamma'_1),\cdots, \mathcal{V}(\Gamma'_r))$, $U_0'=(\mathcal{V}(\Gamma'_{r+1}),\cdots, \mathcal{V}(\Gamma'_{I_1\cdot I_2}))$; $U = (U_1,U_0)$ with $U_1=(\mathcal{V}(\Gamma_1),\cdots, \mathcal{V}(\Gamma_r))$, $U_0=(\mathcal{V}(\Gamma_{r+1}),\cdots, \mathcal{V}(\Gamma_{I_1\cdot I_2}))$; $P\in \mathbb{C}^{r\times r}$, and $\overline{P} \in \mathbb{C}^{(I_1\cdot I_2-r)\times(I_1\cdot I_2-r)}$ are invertible matrices, and $Y$ may be arbitrary.
\label{Lemma-complementary}
\end{lemma}

\noindent{\bf Proof:}  If $\Psi'_{rI_1I_2}$ is SLOCC equivalent to $\Psi_{rI_1I_2}$ then
\begin{eqnarray}
(A_1^{-1} \otimes A_2^{-1}) (\mathcal{V}(\Gamma'_1), \cdots, \mathcal{V}(\Gamma'_{r})) = (\mathcal{V}(\Gamma_1), \cdots, \mathcal{V}(\Gamma_{r}))P \; ,
\end{eqnarray}
where $A_1 \in \mathbb{C}^{I_1\times I_1}$, $A_2 \in \mathbb{C}^{I_2\times I_2}$, and $P\in \mathbb{C}^{r\times r}$ are all invertible matrices. Because the column vectors $(A_1^{-1} \otimes A_2^{-1})\mathcal{V}(\Gamma'_{i})$, $i\in \{r+1,\cdots, I_1\cdot I_2\}$ are linearly independent and belong to the complementary vector spaces of the column vectors of $(\mathcal{V}(\Gamma_1), \cdots, \mathcal{V}(\Gamma_{r}))P$. Therefore there exist an invertible matrix $\overline{P}$ such that
\begin{eqnarray}
(A_1^{-1} \otimes A_2^{-1})(U'_1, U'_0) = (U_1,U_0)
\begin{pmatrix}
  P & Y \\
  0 & \overline{P}
\end{pmatrix} \; .\label{Comp-equiv}
\end{eqnarray}
Here $U'_1=(\mathcal{V}(\Gamma'_1), \cdots, \mathcal{V}(\Gamma'_{r}))$, $U'_0=(\mathcal{V}(\Gamma'_{r+1}), \cdots, \mathcal{V}(\Gamma'_{I_1\cdot I_2}))$, $U_1=(\mathcal{V}(\Gamma_1), \cdots, \mathcal{V}(\Gamma_{r}))$, and $U_0=(\mathcal{V}(\Gamma_{r+1}), \cdots, \mathcal{V}(\Gamma_{I_1\cdot I_2}))$. Therefore $A_1^{-1}\otimes A_2^{-1} = U\widetilde{P}U'^{-1}$, and $\mathrm{rank}[\mathcal{R}( U \widetilde{P} U'^{-1})] = 1$, as the realignment of a matrix has rank one if and only if it is direct product of two other matrices \cite{2LMN,Multi-Criterion}. The converse is quite straightforward. Q.E.D.

\subsection{The SLOCC equivalence of four-partite states}

For two four-partite quantum states $\Psi'$ and $\Psi$ with the triple-state forms of $(\Psi_{u'},\Psi_{\lambda'},\Psi_{v'})$ and $(\Psi_{u},\Psi_{\lambda},\Psi_{v})$, we have the following theorem
\begin{theorem}
Two quadripartite quantum states $\Psi$ and $\Psi'$ are SLOCC equivalent if and only if the states in their corresponding triple-state sets are SLOCC equivalent in the following form
\begin{eqnarray}
|\Psi_{u'}\rangle = P \otimes A_1\otimes A_2\, |\Psi_{u}\rangle \; , \; |\Psi_{v'}\rangle = Q \otimes A_3^* \otimes A_4^*\, |\Psi_{v}\rangle \; , P \otimes Q^{*} \,|\Psi_{\lambda}'\rangle = |\Psi_{\lambda}\rangle \; , \label{Theorem1-eq}
\end{eqnarray}
where $A_1$, $A_2$, $A_3$, $A_4$, $P$, and $Q$ are all invertible matrices. \label{Theorem-four-triple}
\end{theorem}

\noindent{\bf Proof:} First, suppose two four-partite state $\Psi'$ and $\Psi$ are SLOCC equivalent, then $\Psi' = (A_1\otimes A_2) \Psi (A_3\otimes A_4)^{\mathrm{T}}$ and
\begin{eqnarray}
U'\Lambda'V'^{\dag} = (A_1\otimes A_2) U\Lambda V^{\dag} (A_3\otimes A_4)^{\mathrm{T}} \; . \label{ILO-SVD}
\end{eqnarray}
The QR factorizations \cite{Matrix-Analysis} of $(A_1\otimes A_2) U$ and $(A_3^{*}\otimes A_4^{*}) V$ are
\begin{eqnarray}
(A_1\otimes A_2) U = Q_uR_u \; , \; (A_3^{*}\otimes A_4^{*}) V = Q_vR_v\;. \label{QR-factor}
\end{eqnarray}
Here $Q_{u}$ and $Q_{v}$ are unitary matrices, $R_u$ and $R_v$ are upper triangular matrices. Taking Eq. (\ref{QR-factor}) into Eq. (\ref{ILO-SVD}), we have
\begin{eqnarray}
U'\Lambda'V'^{\dag} = Q_uR_u \Lambda R_v^{\dag} Q_v^{\dag}
= Q_uX \Lambda' Y^{\dag}Q_{v}^{\dag} \; ,
\end{eqnarray}
where $R_u\Lambda R_v^{\dag} = X\Lambda' Y^{\dag}$ are the singular value decomposition of $R_u\Lambda R_v^{\dag}$. This leads to the following relations
\begin{eqnarray}
U' = Q_uX(\oplus_i u_{i}) \; , \;  V' = Q_v Y(\oplus u_{i})\; .
\end{eqnarray}
Here $u_i$ are unitary matrices with the dimensions conformal to the degeneracies of the singular values in $\Lambda'$. Considering Eq. (\ref{QR-factor}), we get
\begin{eqnarray}
U' & = & (A_1\otimes A_2) U R_u^{-1} X(\oplus u_i) = (A_1\otimes A_2) U \widetilde{P}\; , \label{Up-U}\\
V' & = & (A_3\otimes A_4)^*  V R_v^{-1} Y(\oplus u_i) = (A_3\otimes A_4)^*  V \widetilde{Q}\; , \label{Vp-V}\\
\Lambda' & = & (\oplus u_i^{\dag})X^{\dag} R_u \Lambda R_v^{\dag}Y (\oplus u_i) = \widetilde{P}^{-1}\Lambda (\widetilde{Q}^{\dag})^{-1} \; , \label{lambdap-lambda}
\end{eqnarray}
where $\widetilde{P} = R_u^{-1} X (\oplus u_i)$ and $\widetilde{Q} = R_v^{-1} Y(\oplus u_i)$; $\Lambda'$ and $\Lambda$ have the following forms
\begin{eqnarray}
\Lambda' = \begin{pmatrix}
\Lambda'_r & 0\\
0 & 0
\end{pmatrix} \; ,\;
\Lambda = \begin{pmatrix}
\Lambda_r & 0\\
0 & 0
\end{pmatrix}\; .
\end{eqnarray}
Here $\Lambda_r'$ and $\Lambda_r$ contain the $r$ nonzero singular values of $\Lambda'$ and $\Lambda$. Therefore the invertible matrices $\widetilde{P}$ and $\widetilde{Q}$ in Eq. (\ref{lambdap-lambda}) take the forms of
\begin{eqnarray}
\widetilde{P} = \begin{pmatrix}
P & Y_1 \\
0 & \overline{P}
\end{pmatrix}\; , \;
\widetilde{Q} =  \begin{pmatrix}
Q & Y_2 \\
0 & \overline{Q}
\end{pmatrix} \; , \label{Th-proof-upper}
\end{eqnarray}
where $P\in \mathbb{C}^{r\times r}$ and $Q \in \mathbb{C}^{r\times r}$ are invertible; $Y_{1,2}$ are submatrices which need not to be identified yet. Based on the representation of Eq. (\ref{Four-Rep}), Eqs. (\ref{Up-U})-(\ref{lambdap-lambda}) lead to Eq. (\ref{Theorem1-eq}).

Second, if $(\Psi_{u'}, \Psi_{\lambda'}, \Psi_{v'})$ and $(\Psi_{u}, \Psi_{\lambda}, \Psi_{v})$ are SLOCC equivalent respectively, i.e. Eq. (\ref{Theorem1-eq}) is satisfied, then by introducing the complementary states and using Lemma \ref{Lemma-complementary}, Eq. (\ref{Comp-equiv}) leads to
\begin{eqnarray}
U' = (A_1\otimes A_2) U \widetilde{P}\; , \; V' = (A_3\otimes A_4)^* V \widetilde{Q} \; , \; \widetilde{P} \Lambda' \widetilde{Q}^{\dag} = \Lambda\; ,
\end{eqnarray}
where $\widetilde{P}$ and $\widetilde{Q}$ have the form of Eq. (\ref{Th-proof-upper}). Therefore
\begin{align}
\Psi' & =  U'\Lambda'V' = (A_1\otimes A_2) U \widetilde{P} \Lambda' \widetilde{Q}^{\dag} V^{\dag} (A_3\otimes A_4)^{\mathrm{T}} \nonumber \\
& = (A_1\otimes A_2) U \Lambda V^{\dag} (A_3\otimes A_4)^{\mathrm{T}} = (A_1\otimes A_2) \Psi (A_3\otimes A_4)^{\mathrm{T}} \; .
\end{align}
That means $\Psi'$ and $\Psi$ are SLOCC equivalent. (Superscripts of transposition on $P$ and $Q$ in Eq. (\ref{Theorem1-eq}) may be needed for consistency, which have no influence on the proof.)
Q.E.D.

Theorem \ref{Theorem-four-triple} turns the SLOCC equivalence of a four-partite state into that of tripartite and bipartite ones. For the SLOCC equivalence of tripartite states, the following Corollary holds.
\begin{corollary}
Two tripartite states $\Psi'_{rI_1I_2}$  and $\Psi_{rI_1I_2}$ are SLOCC equivalent if and only if there exists invertible matrix $\widetilde{P}= \begin{pmatrix}
  P & Y \\
  0 & \overline{P}
\end{pmatrix}$ such that for arbitrary $I_1 \cdot I_2$ vectors $\vec{a}$ we have $\mathrm{rank}[\mathcal{W}(U\widetilde{P}U'^{-1}\vec{a})] = \mathrm{rank}[\mathcal{W}(\vec{a})]$, where $P\in \mathbb{C}^{r\times r}$, $\overline{P} \in \mathbb{C}^{(I_1\cdot I_2-r)\times(I_1\cdot I_2-r)}$ are invertible matrices, and $U$, $U'$ are matrices composed by $\Psi_{rI_1I_2}$, $\Psi'_{rI_1I_2}$ and their complementary states.  \label{Corollary-1}
\end{corollary}

\noindent {\bf Proof:} If $\Psi_{u'}$ and $\Psi_{u}$ are SLOCC equivalent, then according to Eq. (\ref{Comp-equiv}) we have $(A_1^{-1} \otimes A_2^{-1}) U' = U \widetilde{P}$, and
\begin{eqnarray}
\mathcal{W}(U\widetilde{P}U'^{-1}\vec{a}) = \mathcal{W}[(A_1^{-1}\otimes A_2^{-1})\vec{a}\,] = (A_1^{-1})\mathcal{W}(\vec{a}) (A_2^{-1}) \; .
\end{eqnarray}
Hence the ranks of $\mathcal{W}(U\widetilde{P}U'^{-1}\vec{a})$ and $\mathcal{W}(\vec{a})$ are equal for arbitrary $\vec{a}$.

Second, the invertible matrix $\Phi=U\widetilde{P}U'^{-1}$ acting on a vector induces a linear map $\varphi: \mathbb{C}^{I_1\times I_2} \mapsto \mathbb{C}^{I_1\times I_2}$ for the wrapping operation
\begin{eqnarray}
\mathcal{W}(\Phi \vec{a}) = \varphi[\mathcal{W}(\vec{a})]\; .
\end{eqnarray}
Because we have $\mathrm{rank} [\mathcal{W}(\Phi \vec{a})] = \mathrm{rank}[\varphi(\mathcal{W}(\vec{a}))] = \mathrm{rank}[\mathcal{W}(\vec{a})]$ for all $\vec{a}$, the linear map on matrices $\varphi(\mathcal{W}(\vec{a})) = A_1\mathcal{W}(\vec{a})A_2$ follows, where $A_1$ and $A_2$ are invertible matrices according the Theorem 3.1 of Ref. \cite{Linear-Preserver-CKLi}. (Note, when the dimensions $I_1=I_2$, the linear map may be $\varphi(X) = A_1 X^{\mathrm{T}}A_2$, where the two states are SLOCC equivalent up to a permutation of particles.) Q.E.D.

Decomposing a four-partite state into tripartite and bipartite states not only greatly reduces the complexity of the entanglement classification of multipartite states, it also provides a way of studying the multipartite entanglement of the whole system via that of the subsystems. In practice, if we want to verify the SLOCC equivalence of two four-partite states, the equivalence of the two tripartite states should be clarified first.
However Lemma \ref{Lemma-complementary} and Corollary \ref{Corollary-1} provide effective ways for verifying the SLOCC equivalence of tripartite states. It is quite clear that detailed information of the connecting matrices $A_1$, $A_2$, $A_3$, and $A_4$ is not the prerequisite to verify the SLOCC equivalence of arbitrary two four-partite entangled states. We shall show this by the following examples.

\subsection{Examples}

\noindent {\bf Example 1.}  Considering the four-qubit GHZ and W states, i.e., $|\Psi\rangle = \frac{1}{\sqrt{2}}(|0000\rangle + |1111\rangle)$ and $|\Psi'\rangle =  \frac{1}{2}(|0001\rangle + |0010\rangle + |0100\rangle + |1000\rangle)$. According to the partition $(12)(34)$, we have $\Psi_{(12)(34)} = (\Psi_{u},\Psi_{\lambda},\Psi_{v})$ where
\begin{eqnarray}
\Psi_{u} = (\begin{pmatrix}
0 & 0 \\
0 & 1
\end{pmatrix},
\begin{pmatrix}
1 & 0 \\
0 & 0
\end{pmatrix}
)\; ,
\Psi_{\lambda} = \begin{pmatrix}
\frac{1}{\sqrt{2}} & 0 \\
0 & \frac{1}{\sqrt{2}}
\end{pmatrix}\; ,
\Psi_{v} = (\begin{pmatrix}
0 & 0 \\
0 & 1
\end{pmatrix},
\begin{pmatrix}
1 & 0 \\
0 & 0
\end{pmatrix}) \; ,
\end{eqnarray}
and $\Psi'_{(12)(34)} = (\Psi_{u'},\Psi_{\lambda'},\Psi_{v'})$, where
\begin{eqnarray}
\Psi_{u'} =(\begin{pmatrix}
1 & 0 \\
0 & 0
\end{pmatrix},
\begin{pmatrix}
0 & \frac{1}{\sqrt{2}} \\
\frac{1}{\sqrt{2}} & 0
\end{pmatrix})\; ,
\Psi_{\lambda'} = \begin{pmatrix}
\frac{1}{\sqrt{2}} & 0 \\
0 & \frac{1}{\sqrt{2}}
\end{pmatrix}\;
\Psi_{v'}=(\begin{pmatrix}
0 & \frac{1}{\sqrt{2}} \\
\frac{1}{\sqrt{2}} & 0
\end{pmatrix},
\begin{pmatrix}
1 & 0 \\
0 & 0
\end{pmatrix})\; .
\end{eqnarray}
According to Theorem \ref{Theorem-four-triple} and Corollary \ref{Corollary-1}, $\Psi'$ and $\Psi$ are SLOCC equivalent if and only if
\begin{eqnarray}
\mathrm{rank}[ \mathcal{R}(U\widetilde{P}U'^{\dag})] = 1\; , \; \mathrm{rank}[ \mathcal{R}(V\widetilde{Q}V'^{\dag})] = 1\; , \label{Example-GHZW}
\end{eqnarray}
where the submatrices $P$ and $Q$ in $\widetilde{P}$ and $\widetilde{Q}$ should further satisfy $Q=\Psi_{\lambda}^{-1}P\Psi_{\lambda'}$. Eq. (\ref{Example-GHZW}) induces only linear equations on the matrix elements, and we can easily find that $\widetilde{P}=0$ and $\widetilde{Q} = 0$, which indicates GHZ and W states are SLOCC inequivalent.

\noindent {\bf Example 2.} Considering the entangled states $\Psi_{abcd}$ of the first entanglement family in Ref. \cite{four-qubit-nine}, i.e.
\begin{align}
\Psi_{abcd} & = \frac{1}{2}
\begin{pmatrix}
  a+d & 0 & 0 & a-d \\
  0 & b+c & b-c & 0 \\
  0 & b-c & b+c & 0 \\
  a-d & 0 & 0 & a+d
\end{pmatrix} = U\Lambda U^{\dag} \nonumber \\
& = 
\begin{pmatrix}
  \frac{1}{\sqrt{2}} & 0 & 0 & \frac{-1}{\sqrt{2}} \\
  0 & \frac{1}{\sqrt{2}} & \frac{-1}{\sqrt{2}} & 0 \\
  0 & \frac{1}{\sqrt{2}} & \frac{1}{\sqrt{2}} & 0 \\
  \frac{1}{\sqrt{2}} & 0 & 0 & \frac{1}{\sqrt{2}}
\end{pmatrix}
\begin{pmatrix}
  a & 0 & 0 & 0 \\
  0 & b & 0 & 0 \\
  0 & 0 & c & 0 \\
  0 & 0 & 0 & d
\end{pmatrix}
\begin{pmatrix}
  \frac{1}{\sqrt{2}} & 0 & 0 & \frac{-1}{\sqrt{2}} \\
  0 & \frac{1}{\sqrt{2}} & \frac{-1}{\sqrt{2}} & 0 \\
  0 & \frac{1}{\sqrt{2}} & \frac{1}{\sqrt{2}} & 0 \\
  \frac{1}{\sqrt{2}} & 0 & 0 & \frac{1}{\sqrt{2}}
\end{pmatrix} ^{\dag} \; .
\end{align}
Here $V = U$ due to the reason that $\Psi_{abcd}$ is transposition symmetric. Another quantum state $\Psi_{a'b'c'd'} = U\Lambda' U^{\dag}$ is SLOCC equivalent to $\Psi_{abcd}$ if and only if
\begin{eqnarray}
\mathrm{rank}[\mathcal{R}(UPU^{\dag})] = 1\; , \; \mathrm{rank}[ \mathcal{R}(U \Lambda^{-1} P \Lambda' U^{\dag})] = 1 \;, \label{exam-solu}
\end{eqnarray}
where $P$ shall have the same solution in the two equations. $\widetilde{P} = P$ in Eq.(\ref{exam-solu}) because $\Lambda$  has the full rank of 4. As identity matrix is a solution to the first equation of $P$ in Eq. (\ref{exam-solu}),
we get $\Psi_{abcd}$ and $\Psi_{a'b'c'd'}$ are SLOCC equivalent if $\frac{a}{a'} = \frac{b}{b'} = \frac{c}{c'} = \frac{d}{d'}$ or $\frac{a}{a'} = \frac{b}{b'} = \frac{-c}{c'} = \frac{-d}{d'}$ from the second equation in Eq. (\ref{exam-solu}). Other solutions of $P$ would induce more symmetries for the entanglement family parameterized by $a, b, c, d$.

\noindent {\bf Example 3.} Cluster or graph states are highly entangled multiqubit states which are the key resource of measurement base quantum computation \cite{one-way} and various quantum error correction codes \cite{error-corr}. Considering the following four-qubit cluster states
\begin{align}
|C^{(1)}\rangle & = \frac{1}{2}(|0000\rangle + |0101\rangle + |1010\rangle -|1111\rangle) \; , \\
|\Psi^{(2)}\rangle & = a|0000\rangle -b|0111\rangle - c|1010\rangle + d|1101\rangle \; .
\end{align}
Here $|C^{(1)}\rangle $ and $|\Psi^{(2)}\rangle $ are 1-D and 2-D lattice cluster states respectively \cite{1-2-D-Cluster}. Because all the coefficient matrices have the same rank, we do not know whether these two states are SLOCC equivalent or not using the technique of \cite{N-Coefficient-M}. Here we demonstrate the SLOCC equivalence of $|C^{(1)}\rangle $ and $|\Psi^{(2)}\rangle $ based on Theorem \ref{Theorem-four-triple}.

The singular value decomposition according to the bipartition (12)(34) gives
\begin{align}
\Psi^{(2)} = U\Lambda V^{\dag} =
\begin{pmatrix}
1 & 0 & 0 & 0 \\
0 & -1 & 0 & 0 \\
0 & 0 & -1 & 0 \\
0 & 0 & 0 & 1
\end{pmatrix}
\begin{pmatrix}
a & 0 & 0 & 0 \\
0 & b & 0 & 0 \\
0 & 0 & c & 0 \\
0 & 0 & 0 & d
\end{pmatrix}
\begin{pmatrix}
1 & 0 & 0 & 0 \\
0 & 0 & 0 & 1 \\
0 & 0 & 1 & 0 \\
0 & 1 & 0 & 0
\end{pmatrix} \; , \\
C^{(1)} = U'\Lambda' V'^{\dag} =
\begin{pmatrix}
1 & 0 & 0 & 0 \\
0 & 1 & 0 & 0 \\
0 & 0 & 1 & 0 \\
0 & 0 & 0 & -1
\end{pmatrix}
\begin{pmatrix}
\frac{1}{2} & 0 & 0 & 0 \\
0 & \frac{1}{2} & 0 & 0 \\
0 & 0 & \frac{1}{2} & 0 \\
0 & 0 & 0 & \frac{1}{2}
\end{pmatrix}
\begin{pmatrix}
1 & 0 & 0 & 0 \\
0 & 1 & 0 & 0 \\
0 & 0 & 1 & 0 \\
0 & 0 & 0 & 1
\end{pmatrix} \; ,
\end{align}
from which we may get $(\Psi_{u},\Psi_{\lambda}, \Psi_{v})$ and $(\Psi_{u'},\Psi_{\lambda'}, \Psi_{v'})$. The SLOCC equivalence of the resulted $\Psi_{u}$ and $\Psi_{u'}$, and $\Psi_{v}$ and $\Psi_{v'}$ may be verified by exploring Lemma \ref{Lemma-complementary}. Similarly as Eq. (\ref{exam-solu}) we have that the following two matrices must have rank 1
\begin{align}
\mathcal{R}(U P U'^{\dag})& = \begin{pmatrix}
  p_{11} & -p_{21} & p_{12} & -p_{22} \\
  -p_{31} & p_{41} & -p_{32} & p_{42} \\
  p_{13} & -p_{23} & -p_{14} & p_{24} \\
  -p_{33} & p_{43} & p_{34} & -p_{44}
\end{pmatrix} \; , \label{UPUp} \\
\mathcal{R}(V\Lambda^{-1}P \Lambda' V'^{\dag}) & = \frac{1}{2} \begin{pmatrix}
  \frac{p_{11}}{a} & \frac{p_{41}}{d} & \frac{p_{12}}{a} & \frac{p_{42}}{d} \\
  \frac{p_{31}}{c} & \frac{p_{21}}{b} & \frac{p_{32}}{c} & \frac{p_{22}}{b} \\
  \frac{p_{13}}{a} & \frac{p_{43}}{d} & \frac{p_{14}}{a} & \frac{p_{44}}{d} \\
  \frac{p_{33}}{c} & \frac{p_{23}}{b} & \frac{p_{34}}{c} & \frac{p_{24}}{b}
\end{pmatrix} \; . \label{VPVp}
\end{align}
Here $p_{ij}$ are the matrix elements of $P$ and we have used the relation $ V\Lambda^{-1}P\Lambda' V'^{\dag} = (VQV'^{\dag})^{-1\dag}$. The solutions for Eq. (\ref{UPUp}) and Eq. (\ref{VPVp}) having rank 1 are
\begin{equation}
P = \begin{pmatrix}
 p_{11} & p_{12} & x p_{11} & -x p_{12} \\
 p_{21} & p_{22} & x p_{21} & -x p_{22} \\
 -y p_{11} & -y p_{12} & -z p_{11} & z p_{12} \\
 -y p_{21} & -y p_{22} & -z p_{21} & z p_{22} \\
\end{pmatrix} \; ,\;
P = \begin{pmatrix}
 p_{11} & p_{12} & \alpha p_{11} & \alpha p_{12} \\
 p_{21} & p_{22} & \frac{\gamma}{\beta} p_{21} & \frac{\gamma}{\beta} p_{22} \\
\frac{c\beta}{a} p_{11} & \frac{c\beta}{a} p_{12} & \frac{c\gamma}{a} p_{11} & \frac{c\gamma}{a} p_{12} \\
\frac{d}{b\beta} p_{21} & \frac{d}{b\beta} p_{22} & \frac{d\alpha}{b\beta} p_{21} & \frac{d\alpha}{b\beta} p_{22} \\
\end{pmatrix} \; , \label{P-sol-Q}
\end{equation}
where $x,y,z$ and $\alpha,\beta,\gamma$ are nonzero parameters. A consistent solution of matrix $P$ is
\begin{equation}
p_{12}=p_{21}=0\;, \; x=\alpha=-\frac{\gamma}{\beta}\; , \; y = -\frac{c\beta}{a} = -\frac{d}{b\beta}\; , \;z = -\frac{c\gamma}{a} = \frac{d\alpha}{b\beta} \; . \label{Pvalue}
\end{equation}
Eq. (\ref{Pvalue}) predicts an invertible matrix $P$, and therefore $|C^{(1)}\rangle$ and $|\Psi^{(2)}\rangle$ are SLOCC equivalent. After getting $P$, we may also easily get
\begin{equation}
A_1 =
\frac{1}{2} \begin{pmatrix}
1 & \frac{1}{y} \\
\frac{1}{x} & \frac{1}{z}
\end{pmatrix} \; , \;
A_2 = \begin{pmatrix}
\frac{1}{p_{11}} & 0 \\
0 & \frac{-1}{p_{22}}
\end{pmatrix} \; , \;
A_3 =
\frac{1}{2} \begin{pmatrix}
1 & \beta \\
\alpha & \gamma
\end{pmatrix} \; , \;
A_4 = \begin{pmatrix}
\frac{p_{11}}{a} & 0 \\
0 & \frac{p_{22}}{b\beta}
\end{pmatrix} \; ,
\end{equation}
which connect the matrices $|C^{(1)}\rangle = A_1\otimes A_2\otimes A_3\otimes A_4 |\Psi^{(2)}\rangle$. The SLOCC equivalence of $|C^{(1)}\rangle$ and $|\Psi^{(2)}\rangle$ indicates that different dimensional (1-D and 2-D) cluster states may be equivalent in realizing quantum computation tasks, which is important to the study of measurement-based quantum computation models.

The above examples indicate that the new method works effectively for finite dimensional four-partite systems. Different choices of partitions, i.e. (13)(24) and (12)(34) for four-partite states, do not influence the application of the method. However, one must choose the same partition for two four-partite states when verifying their SLOCC equivalence. It should be noted that the entangled quantum states may be of infinite dimensional \cite{RMD-Continuous}. Although it is still unclear whether our method is applicable to this situation or not, the reduction method is nevertheless inspiring to the study of continuous-variable entanglement.

\section{Conclusions}

We proposed a practical classification scheme for four-partite entangled state, in which the introduced neat mathematical trick defines a ¡®virtual system¡¯ with subsystems different from the original one, whose entangled structure however faithfully represents the SLOCC relations of the original system. According to this scheme, a prerequisite for connecting matrices between two four-partite states is unnecessary, which greatly reduces the complexity in usual procedure in verifying the SLOCC equivalence. According to the reduction method of this work, the relations between a high partite entangled state, the four-partite state in this work, and its subsystems and bridges between them turn out to be manifest. It is notable that the method developed here opens the door of hopes to the the general multipartite entanglement SLOCC classification. Furthermore, it is tempt to think that the high order singular value decomposition technique in the local unitary(LU) classification of multipartite entangled state is worthy of our reference \cite{LU-HOSVD}.
\\

\section*{Acknowledgements}

\noindent
This work was supported in part by the Ministry of Science and Technology of the People¡¯s Republic of China(2015CB856703); by the Strategic Priority Research Program of the Chinese Academy of Sciences, Grant No.XDB23030100; and by the National Natural Science Foundation of China(NSFC) under the grants 11375200 and 11635009. S.M.Z. is also supported in part by the CAS-TWAS fellowship.

\end{document}